\documentclass[prd,showpacs,amsmath,amssymb]{revtex4-2}
\usepackage{graphicx}
\usepackage{dcolumn}
\usepackage{bm}
\usepackage{besphysics}
\usepackage{besrefblk}
\usepackage{comment}
\usepackage{multirow}
\usepackage{ulem}
\usepackage{amsmath}
\usepackage[colorlinks,
linkcolor=blue,
anchorcolor=blue,
citecolor=blue
]{hyperref}
\usepackage{rotating}
\usepackage{color}
\usepackage{xcolor}

\newcommand{\zcsn}{Z_{cs}^0}

\newcommand{\DDsone}{\Dsp\Dstm}
\newcommand{\DDstwo}{\Dsstp \Dm}
\newcommand{\br}[1]{\mathcal{B}(#1)}

\renewcommand{\thefootnote}{\fnsymbol{footnote}}
\begin{document}
\normalsize
\parskip=5pt plus 1pt minus 1pt
\title{\boldmath Evidence for a neutral near-threshold structure in the $K^{0}_{\rm S}$ recoil-mass spectra
in $e^+e^-\rightarrow K^{0}_{\rm S}D_s^+D^{*-}$ and $e^+e^-\rightarrow K^{0}_{\rm S}D_s^{*+}D^{-}$}
\author{\small M.~Ablikim$^{1}$, M.~N.~Achasov$^{11,b}$, P.~Adlarson$^{70}$, M.~Albrecht$^{4}$, R.~Aliberti$^{31}$, A.~Amoroso$^{69A,69C}$, M.~R.~An$^{35}$, Q.~An$^{66,53}$, X.~H.~Bai$^{61}$, Y.~Bai$^{52}$
% ...
\\
\vspace{0.2cm}
(BESIII Collaboration)\\
\vspace{0.2cm} {\it
$^{1}$ Institute of High Energy Physics, Beijing 100049, People's Republic of China\\
% ...
$^{4}$ Bochum Ruhr-University, D-44780 Bochum, Germany\\
% ...
$^{11}$ G.I. Budker Institute of Nuclear Physics SB RAS (BINP), Novosibirsk 630090, Russia\\
% ...
$^{31}$ Johannes Gutenberg University of Mainz, Johann-Joachim-Becher-Weg 45, D-55099 Mainz, Germany\\
% ...
$^{35}$ Liaoning Normal University, Dalian 116029, People's Republic of China\\
% ...
$^{52}$ Southeast University, Nanjing 211100, People's Republic of China\\
$^{53}$ State Key Laboratory of Particle Detection and Electronics, Beijing 100049, Hefei 230026, People's Republic of China\\
% ...
$^{61}$ University of Jinan, Jinan 250022, People's Republic of China\\
% ...
$^{66}$ University of Science and Technology of China, Hefei 230026, People's Republic of China\\
% ...
$^{69}$ University of Turin and INFN, (A)University of Turin, I-10125, Turin, Italy; (B)University of Eastern Piedmont, I-15121, Alessandria, Italy; (C)INFN, I-10125, Turin, Italy\\
$^{70}$ Uppsala University, Box 516, SE-75120 Uppsala, Sweden\\
% ...
$^{b}$ Also at the Novosibirsk State University, Novosibirsk, 630090, Russia\\
% ...
}
}
\vspace{0.4cm}
\begin{abstract}
We study the processes $e^+e^-\rightarrow K^{0}_{\rm S}D_s^+D^{*-}$ and $e^+e^-\rightarrow K^{0}_{\rm S}D_s^{*+}D^{-}$, as well as their charge conjugated processes,
at five center-of-mass energies between 4.628~GeV and 4.699~GeV, using data samples corresponding to an integrated luminosity of $3.8~\rm{fb^{-1}}$ collected by the BESIII detector
at the BEPCII storage ring. Based on a partial reconstruction technique, we find evidence of a
structure near the thresholds for $D_s^+D^{*-}$ and $D_s^{*+}D^{-}$ production in the
$K^{0}_{\rm S}$ recoil-mass spectrum, which we refer to as the $ Z_{cs}(3985)^0$.
Fitting with a Breit-Wigner line shape, we find the mass of the structure to be $(3992.2\pm{1.7}\pm{1.6})~\rm{MeV}/c^2$ and
the width to be  $(7.7_{-3.8}^{+4.1}\pm{4.3})~\rm{MeV}$, where the first uncertainties are
statistical and the second are systematic. The significance of the
$ Z_{cs}(3985)^0$ signal is found to be $4.6\sigma$ including both the statistical and systematic uncertainty. We report the Born cross section multiplied by the branching
fraction at different energy points. 
The mass of the $Z_{cs}(3985)^0$ is close to that of the $Z_{cs}(3985)^+$. Assuming SU(3) symmetry, the cross section of the neutral channel is consistent with that of the charged one. Hence, we conclude that the $Z_{cs}(3985)^0$ is the isospin partner of the $Z_{cs}(3985)^+$. 
\end{abstract}

\maketitle

Extensive evidence exists for several non-strange hidden-charm tetraquark
$Z_c$ candidates, with quark constituent of $c\bar{c}q\bar{q}'$ ($q^{(\prime)}=u$
or $d$)~\cite{XYZ_review1,XYZ_review2,XYZ_review3,XYZ_review4}. 
In electron-positron annihilation, both charged and neutral $Z_c(3900)$
and $Z_c(4020)$ states have been observed by the BESIII, Belle and CLEO collaborations
\cite{Ablikim:2013mio1,Ablikim:2013wzq1,Ablikim:2013xfr1,Ablikim:2013emm1,Ablikim:2014dxl,Ablikim:2015tbp,Ablikim:2015vvn,Ablikim:2015gda,BESIII:2015pqw,Liu:2013dau,Xiao:2013iha}.
Under SU(3) flavor symmetry, one expects the existence of corresponding strange partners
with $c\bar{c}s\bar{q}$  configurations, denoted as $ Z_{cs}$ states~\cite{Voloshin:2019ilw}.
These $ Z_{cs}$ states are predicted to have masses close to the $D_s\bar{D}^*$ and $D_s^*\bar{D}$ thresholds in a variety models explaining their nature, including the
tetraquark scenario~\cite{Lee:2008uy, Ferretti:2020ewe}, 
the molecular model~\cite{Dias:2013qga}, the hadron-quarkonium
model~\cite{Ferretti:2020ewe}, and the initial-single-chiral-particle-emission
mechanism~\cite{Chen:2013wca}.

The charged-tetraquark candidate
$Z_{cs}(3985)^+$~\cite{Zcs} was observed at BESIII in the $\Dsp\bar{D}^{*0}$ and
$\Dsstp\bar{D}^{0}$ final states
~\cite{Meng:2020ihj,Wan:2020oxt,Yang:2020nrt,Maiani:2021tri,Meng:2021rdg}. The mass of the $Z_{cs}(3985)^+$ is close to the $\Dsp \bar{D}^{*0}$ and $\Dsstp \bar{D}^{0}$ thresholds, which is consistent with theoretical predictions~\cite{Lee:2008uy,Ferretti:2020ewe,Dias:2013qga,Chen:2013wca}. Meanwhile, another charged-tetraquark candidate, $Z_{cs}(4000)^+$~\cite{LHCb}, was observed in the $J/\psi K^{+}$ final states in an amplitude analysis of the decay $B^{+}\rightarrow J/\psi\phi K^{+}$ at LHCb.  However, the widths of these two $Z_{cs}^+$ states are inconsistent with each other.
The observation of these charged $Z_{cs}$ states motivates a search for a neutral isospin partner $\zcsn$.
The mass of the $\zcsn$ is expected to be heavier than that of the $Z_{cs}^+$ by $(0.05\pm0.21)$ GeV/$c^2$
under the molecular hypothesis, or by $(0.06\pm0.12)$ GeV/$c^2$
under the tetraquark hypothesis~\cite{Wan:2020oxt}. 
A promising approach to this challenge at BESIII is to search for the process  $\ee\to\bar{K}^0 Z_{cs}(3985)^0+c.c.$ and then compare its cross section to that of $\ee\to K^-Z_{cs}(3985)^++c.c.$, which tests the isospin symmetry in the production and decay dynamics.  A similar strategy was pursued in the analysis of the $Z_c$ charged and neutral states~\cite{Ablikim:2015tbp,Ablikim:2015vvn}.   Observation and study of the $\zcsn$ is crucial for understanding the nature of the $Z_{cs}$ states.

In this letter, we study the processes $e^+e^-\rightarrow K^{0}_{\rm S}D_s^+D^{*-}$ and $e^+e^-\rightarrow K^{0}_{\rm S}D_s^{*+}D^{-}$, which is denoted as $\ee\to\kshort(\DDsone +
\DDstwo)$ in the context, as well as their charge conjugated modes, using $e^+e^-$ collision data sets corresponding to an integrated luminosity of $3.8~\rm{fb^{-1}}$~\cite{lum} at center-of-mass energies $\sqrt{s}= 4.628, 4.641, 4.661, 4.682$ and $4.699\gev$~\cite{lum}. These samples were collected by the BESIII detector at the Beijing Electron Positron Collider (BEPCII). 
Detailed information about BEPCII and BESIII can be found in Refs.~\cite{Yu:IPAC2016-TUYA01, Ablikim:2009aa, Ablikim:2019hff}.
We use a partial reconstruction technique to maximize the detection efficiency; only the $\kshort$ produced in association with the $D_s^+D^{*-}$ or $D_s^{*+}D^{-}$ (the \textit{bachelor} $\kshort$) and one of the ground-state $D$ mesons (here $D$ subsequently denotes $\Dsp$ or $\Dm$) are detected, while the other final-state particles are not reconstructed. The $Z_{cs}^0$ candidate is then searched for in the invariant mass distribution  recoiling against the bachelor $\kshort$ candidate. Charge conjugation is implied throughout the discussion.

Simulated samples produced with a {\sc geant4}-based~\cite{geant4} Monte Carlo
(MC) package, which includes the geometric description of the BESIII detector and
the detector response, are used to determine the detection efficiency and to
understand the backgrounds. The $e^+e^-$ annihilations are simulated with the
{\sc kkmc}~\cite{ref:kkmc} generator, which includes the effects of the beam-energy spread and initial-state radiation (ISR). The inclusive MC sample consists of the
production of open-charm hadronic systems, ISR production of vector
charmonium(-like) states, and continuum processes incorporated in {\sc
kkmc}~\cite{ref:kkmc}. The known decay modes are modelled with {\sc
evtgen}~\cite{ref:evtgen} using branching fractions reported by the Particle Data
Group (PDG)~\cite{pdg}, and the remaining unknown decays from  charmonium states are modelled
with {\sc lundcharm}~\cite{ref:lundcharm}. The final-state radiation (FSR) from
charged final-state particles is simulated with the {\sc photos}
package~\cite{photos}.
For the non-resonant three-body signal processes $\ee\to\kshort(\Dsp\Dstm + \Dsstp\Dm)$, the momenta distributions of final-state particles are generated following phase space. For the resonant signal process
$\ee\to\kshort\zcsn\to \kshort(\Dsp\Dstm+\Dsstp\Dm)$, we assume that the $\zcsn$ state has a spin-parity of $1^{+}$, which corresponds to $S$-waves in both of the decays  $\ee\to\kshort\zcsn$ and
$\zcsn\to\Dsp\Dstm+\Dsstp\Dm$, which we denote as ($S$, $S$). The corresponding angular
distribution is taken into account in simulating the cascade decays. 
Other possibilities for the $Z_{cs}^0$ spin-parity are tested to evaluate the systematic uncertainty related to this assumption. \par

We carry out two types of partial
reconstruction, which are referred as the $\Dsp$-tag and $\Dm$-tag methods, respectively. For 
$\Dsp$($\Dm$)-tag method, only the bachelor $\kshort$ and $\Dsp$($\Dm$) candidates are reconstructed. 
We use the decay modes $\Dsp \to
\kaonp\kaonm\pip$, $\kshort\kaonp$, $\kaonp\kaonm\pip\piz$,
$\kshort\kaonp \pip\pim$ and $\eta'\pip$ to form the $\Dsp$ candidates; and the decay modes $\Dm \to \kaonp\pim\pim$, $\kshort\pim$ and $\kshort \pip \pim \pim$ to form the $\Dm$ candidates.

To ensure that each charged track, which is not associated to $\kshort$ detection, originates 
from the $e^+e^-$ interaction point (IP), $|V_r|<1$~cm
and $|V_z|<10$~cm are required. Here, $|V_r|$ is the distance between the charged track 
and the beam axis in the transverse plane, and $|V_z|$ is the closest distance of the charged track to the IP along  the axis of beam. The polar angles of 
charged tracks are required to satisfy $|\rm cos\theta|<0.93$.
The flight time in the time-of-flight system
and the energy deposited in the multilayer drift
chamber 
for each charged track are used to identify particles by
calculating the probabilities $\rm{P}(i)$, where $i$ denotes $K$ or $\pi$. We require ${\rm P}(K)$ (${\rm P}(\pi)$) to be greater than ${\rm P}(\pi$) (${\rm P}(K)$) to classify a particle as a kaon (pion) candidate.

The $\kshort$ candidates are reconstructed through the $\pip\pim$ decay mode without particle identification  requirements.
Both pions must satisfy $|V_z|<20$~cm, and 
$|\rm cos\theta|<0.93$ and their trajectories are constrained to originate from a common vertex by applying a vertex fit, the $\chi^2$ of which is required to be less than 100. The $\kshort$ candidate is then formed and the opposite direction of its momentum is constrained to point at the IP, with the corresponding $\chi^2$ required to be less than 40.
The decay length of $\kshort$ candidate must be greater than two standard deviations of the vertex
resolution away from the IP. The invariant mass of $\pip\pim$ pair, $M(\pip\pim)$, is
required to be within $(0.492,0.503)~{\rm GeV}/c^2$. 

The $\piz$ and $\eta$ candidates are reconstructed through $\piz/\eta\rightarrow\gamma\gamma$.  The photon showers in the electromagnetic calorimeter must have energies greater than 25 MeV in the barrel region
($\left|\cos\theta\right|<0.80$) and greater than 50~MeV in the end-cap region
($0.86<|\cos\theta|<0.92$). Showers must have an associated time within 700~ns of the event start time. A kinematic fit is applied to constrain the invariant mass of the $\gamma\gamma$ pair to the known $\piz$ or $\eta$ mass reported in the PDG~\cite{pdg}, and the resultant $\chi^2$ is required to be less than 10. 
The $\eta'$ is reconstructed through $\eta'\rightarrow\pip\pim\eta$. The mass of the $\pip\pim\eta$ is required to be within 10~MeV/$c^2$ of the known $\eta'$~\cite{pdg} mass.

To improve the signal purity, the requirements listed in Table~\ref{tab:cuts} are adopted to restrict the final states within the regions of the $\phi$, $K^*$ and $\rho$ resonances, which dominate the decays.
In the selection of $\Dm\rightarrow \kshort\pip\pim\pim$
candidates, contamination from the decay $\Dm\rightarrow \kshort \kshort\pim$ is suppressed by
requiring the invariant mass of the $\pip\pim$ pair to lie outside the interval $(0.48,0.52)~{\rm GeV}/c^2$.
To avoid double counting and to suppress backgrounds, we only keep the $\Dsp$ ($\Dm$) candidate with an invariant mass closest to the known $\Dsp$ ($\Dm$) mass.
In the invariant-mass spectra of the $D$ decay final states, the signal candidates are selected by requiring  the reconstructed mass $M(D)$  to be within $15~{\rm MeV}/c^2$ of the known mass of the charm meson in question.
The $M(\Dsp)$ sideband regions are defined as (1.895, 1.935) $~{\rm GeV}/c^2$ and (1.995, 2.035) $~{\rm GeV}/c^2$, while
the $M(\Dm)$ sideband regions are defined as (1.800, 1.840) $~{\rm GeV}/c^2$ and (1.900, 1.940) $~{\rm GeV}/c^2$, which are taken as control samples for studying  the combinatorial backgrounds in the subsequent analysis.

\begin{figure}
\centering
\includegraphics[width=0.48\textwidth]{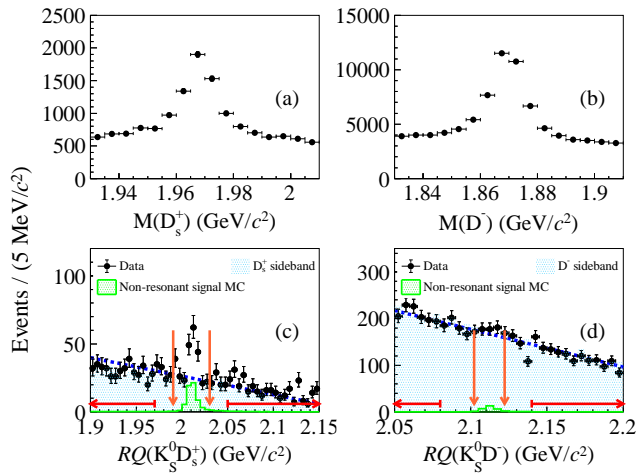}
\caption{Invariant-mass distributions of the singly tagged $\Dsp$ (a) and $\Dm$ (b), together with the fits to the recoil-mass distributions $RQ(\kshort \Dsp)$ (c) and $RQ(\kshort\Dm)$ (d) at 4.682 GeV.
The points with error bars are data. The blue-dashed lines show the fit results of the sideband regions, which are denoted by the red arrows. The histograms show the distributions from the non-resonant signal MC samples, which are scaled according to the yields of $D^{*-}$ and $D_{s}^{*+}$. The orange arrows indicate the signal regions of the $D^{*-}$ in (c) and $D_{s}^{*+}$ in (d).}
\label{fig:M_recKsDs}
\end{figure}

\begin{table}[tp]
\centering
\caption{Summary of the cuts applied to the $\Dsp$ and $\Dm$ decay modes for the combinatorial background suppression. Here $M$ denotes the reconstructed invariant mass and $m$ the known mass.}
\renewcommand{\arraystretch}{1.3}
\begin{tabular}{cc} \hline\hline 
Final state & Requirement \\
\hline                                         
\multirow{2}{*}{$\Dsp\rightarrow \kaonp\kaonm\pip$}         & $M(\kaonp\kaonm)<1.05~{\rm GeV}/c^2$              \\
                                       & $|M(\kaonp\pim)-m(K^*(892))|<70~{\rm MeV}/c^2$    \\
\hline                                           
\multirow{2}{*}{$\Dsp\rightarrow \kaonp\kaonm\pip\piz$}    & $M(\kaonp\kaonm)<1.05~{\rm GeV}/c^2$              \\
                                       & $|M(\pim\piz)-m(\rho)|<150~{\rm MeV}/c^2$    \\
\hline                                           
$\Dsp\rightarrow \kshort\kaonp\pip\pim$  & $|M(\kaonp\pim)-m(K^*(892))|<70~{\rm MeV}/c^2$    \\
\hline                                           
$\Dm\rightarrow \kshort\pip\pim\pim$  & $|M(\kshort\pip)-m(K^*(892))|<70~{\rm MeV}/c^2$    \\
\hline\hline
\end{tabular}
\label{tab:cuts}
\end{table}

The recoil mass $RM(\kshort D)$ of the $\kshort D$ system is obtained according to $RM(X)=||p_{e^+e^-}-p_X||$, where $p_{e^+e^-}$ is the four-momentum of the initial $e^+e^-$ system and $p_X$ is the four-momentum of the $X$ system. 
The $RM(\kshort D)$ resolution is then improved through use of the quantity $RQ(\kshort D)=RM(\kshort D)+M(D)-m(D)$~\cite{Ablikim:2018qjv}, where  $M(D)$ is the invariant mass of the signal $D$ candidate, and $m(D)$ is the known mass quoted in PDG~\cite{pdg}.  The $RQ(\kshort D)$ spectra are shown in Fig.~\ref{fig:M_recKsDs}. 
These spectra are used
to identify the three body processes $\kshort\Dsp\Dstm$ and
$\kshort\Dsstp\Dm$, which contribute to peaking structures in the  regions of the $\Dstm$ and $\Dsstp$ mass, respectively. 
We require  $|RQ(\kshort\Dsp)-m(\Dstm)|<20\mevcc$ and $|RQ(\kshort\Dm)-m(\Dsstp)|<10\mevcc$.
Studies of the inclusive MC simulations show that there is negligible peaking background in the signal regions.
To evaluate the level of combinatorial background in the sample of selected three-body candidates, linear fits to the $RQ(\kshort\Dsp)$ sideband region ($[1.90, 1.97]\gevcc$ and
$[2.05, 2.15]\gevcc$), and to the $RQ(\kshort\Dm)$ sideband region ($[2.05, 2.08]\gevcc$ and $[2.14, 2.20]\gevcc$) are performed, where the slopes are fixed according to the corresponding $M(D)$ sideband samples.
Table~\ref{tab:combbkg} lists the number of combinatorial background candidates for the two tag methods at each energy point.

\begin{table}[tp]
\centering
\caption{Number of combinatorial background candidates in the signal regions of the $\kshort\Dsp\Dstm$ and $\kshort\Dsstp\Dm$
three-body processes.}
\begin{tabular}{c r@{$\pm$}l r@{$\pm$}l} \hline\hline
$\sqrt{s}$(MeV)  &  \multicolumn{2}{c}{$\Dsp$-tag} & \multicolumn{2}{c}{$\Dm$-tag}         \\
\hline                                           
4628 & 40.6&3.4& 132.1&6.1 \\
4641 & 49.8&3.7& 169.1&6.8 \\
4661 & 57.5&4.0& 184.3&6.9 \\
4682 &199.0&7.3& 668.8&12.9 \\
4699 & 68.6&4.2& 217.5&7.4 \\
\hline\hline
\end{tabular}
\label{tab:combbkg}
\end{table}

\begin{figure}[tp]
\centering
\includegraphics[width=0.48\textwidth]{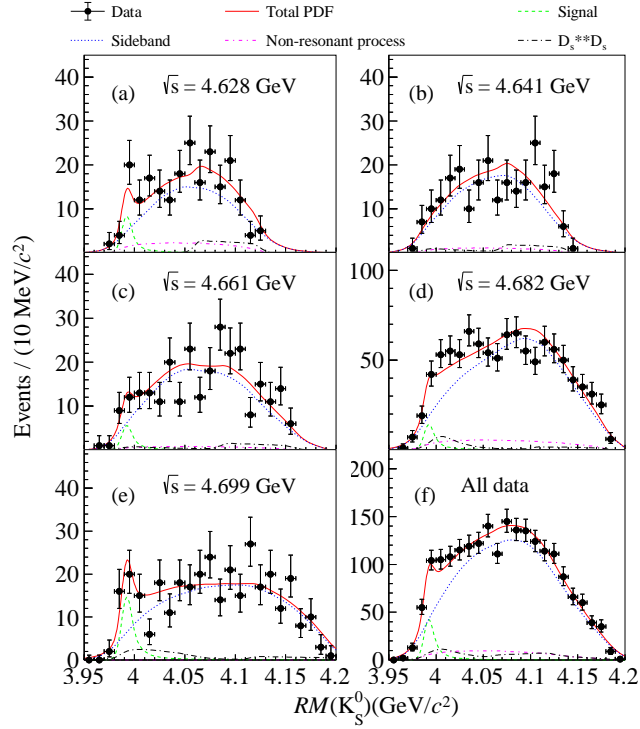}
\caption{Simultaneous fit to the recoil mass $RM(\kshort)$ spectra in all data sets [(a) to (e)], and for all the data points combined (f).  The green dashed curves show the $Z^0_{cs}$ signal contribution. The pink
dash-dotted curves show the non-resonant process. The blue dotted curves show combinatorial
backgrounds. The black long dash-dotted curves show contributions from the highly excited
$D^{**}_{(s)}$ backgrounds.}
\label{fig:fit_withZcs}
\end{figure}

Fig.~\ref{fig:fit_withZcs} shows $RM(K_{\rm S}^0)$, the bachelor $K_{\rm S}^0$ recoil-mass distribution,  for the signal candidates selected from both tags. 
There is an enhancement near the mass threshold of $\Dsp\Dstm$ and $\Dsstp\Dm$, which is most evident in the  $4.682\gev$ and $4.699\gev$ data sets. 

To understand potential contributions from the highly excited strange-charmed mesons $D^{**}_{s}$ in the $RM(\kshort)$ distribution, we simulate the exclusive production of $D_{s1}(2536)^{-}D_{s}^{+}$, $D_{s2}(2573)^{-}\Dsstp$, and
$D_{s1}(2700)^{-}D_{s}^{+}$ in $\ee$ annihilations. Assuming isospin symmetry, their production cross sections are those of the corresponding states studied during the analysis of the charged
$ Z_{cs}(3985)^+$\cite{Zcs}. In addition, the potential effect of excited non-strange charmed mesons $D^{**}$ are explored as described in Ref.~\cite{supple}, where $D_2^*(2460)^+\Dstm$, $D(2550)^+D^{-}$, $D_1^*(2600)^+\Dstm$, $D_1^*(2600)^+D^{-}$, $D(2740)^+D^{-}$, and $D_3^*(2750)^+\Dm$ are taken into account.
We find the threshold enhancement can not be explained by these excited states, and hence, we consider its possible origin to be the neutral $Z_{cs}^0$ state.

A simultaneous unbinned maximum likelihood fit is applied to the distributions of $RM(\kshort)$
at five energy points. 
We adopt two $S-$wave
Breit-Wigner functions $R_1$ and $R_2$ to describe the $Z^0_{cs}$ resonance
\begin{small}
$$R=\left|\frac{1}{M^2-m^2_0+im_0(f\cdot\Gamma_1(M)+(1-f)\cdot\Gamma_2(M))}\right|^{2}$$
$$R_1=R\cdot
q\cdot p_1,$$
$$R_2=R\cdot
q\cdot p_2,$$
$$\Gamma_1(M)=\Gamma_0\cdot\frac{p_1}{p^{*}_1}\cdot\frac{m_0}{M},$$
$$\Gamma_2(M)=\Gamma_0\cdot\frac{p_2}{p^{*}_2}\cdot\frac{m_0}{M},$$\\
\end{small}
where $R_1$
describes the decay $Z^0_{cs}\rightarrow \Dsp\Dstm$, and $R_2$ describes $Z^0_{cs}\rightarrow\Dsstp\Dm$, $M$ equals $RM(\kshort)$, $m_0$ is the mass of the $Z_{cs}^0$, and $\Gamma_0$ is the total width of the $Z_{cs}^0$. The momentum of the $\kshort$ in the initial
$e^+e^-$ system is $q$, the momentum of the $\Dsp(\Dm)$ in the rest frame
of the $\Dsp\Dstm(\Dsstp\Dm)$ system is $p_{1(2)}$, and the corresponding momentum at $M=m_{0}$ is $p^{*}_{1(2)}$. In the fit, under the assumption of the isospin symmetry,
a Gaussian constraint is imposed to restrict the width of the $\zcsn$  
within the uncertainty of the $Z_{cs}(3985)^+$ width, which is $(13.8_{-5.2}^{+8.1}\pm{4.9})~\rm{MeV}$\cite{Zcs}. The
factor $f$ denotes the ratio of the two signal channels
\begin{equation}
f=\frac{\br{Z^0_{cs}\rightarrow \Dsp\Dstm}}{\br{Z^0_{cs}\rightarrow \Dsp\Dstm}+\br{Z^0_{cs}\rightarrow\Dsstp\Dm}}.
\label{eq:fFactor}
\end{equation}
The default value of $f$ is chosen to be 0.5, with other possibilities considered as a systematic uncertainty.

The fit depends on the detector resolution and mass-dependent efficiency, which are derived from simulated samples. The detector resolution is determined using the $Z_{cs}^{0}$ signal MC samples, in which the width of the
$Z_{cs}^{0}$ is set to be 0.
The signal probability density function (PDF) is constructed as follows:
\begin{equation}
\mathcal{F}\propto(f\cdot\mathcal{E}_{1}\cdot R_1+(1-f)\cdot\mathcal{E}_{2}\cdot R_2)\otimes G(\mu,\sigma), 
\end{equation}
where $\mathcal{E}_{1(2)}$ is the efficiency function and $G$ is the Gaussian resolution function.

The backgrounds in the fit include three components: the non-resonant process 
$e^+e^-\rightarrow \kshort(\Dsp\Dstm+\Dsstp\Dm)$, the excited
$D^{**}_{s}D_{s}$ backgrounds, and the combinatorial backgrounds. The first and
second components are described using histogram PDFs extracted from MC
samples, and the third component is described using the distribution from
the $\Dsp$ $(\Dm)$ sideband. 
In the fit, the yields of the excited $D^{**}_{s}D_{s}$ backgrounds are
estimated from isospin relations according to those calculated for the $e^+e^-\rightarrow \kaonm Z_{cs}(3985)^+$ process, and the
numbers are fixed in the fit~\cite{Zcs}, while the yields of the non-resonant process are free. The sizes of the combinatorial background are
fixed to the values in Table~\ref{tab:combbkg}.

The fitted mass and width of the $Z_{cs}^0$ are given in 
Table~\ref{tab:CompareZcs}, where the $Z_{cs}(3985)^+$ resonance parameters are included for comparison.
The results are consistent with the theoretical predictions~\cite{Lee:2008uy,Ferretti:2020ewe,Dias:2013qga,Chen:2013wca,Wan:2020oxt}.
We sum up the  $RM(\kshort)$ distributions from all data sets, and superimpose the simultaneous fit curves in the last plot of Fig.~\ref{fig:fit_withZcs}.
Comparing the fits with or without considering the contribution from the
$Z_{cs}^0$, 
the number of degrees of freedom is changed by 7 (the mass and
width of the $Z_{cs}^0$, together with the cross section of the $Z_{cs}^0$ at the five
center-of-mass energies). The value of $2\ln{L}$, where $L$ is the likelihood value, is changed by 42.0. This
corresponds to a statistical significance of $5.0\sigma$ according to Wilks' theorem~\cite{Wilk}. When also considering systematic
uncertainties, which are described in the supplemental material~\cite{supple}, the significance of the $Z_{cs}^0$ signal becomes $4.6\sigma$. The reduced chi-squared of the fit in Fig.~\ref{fig:fit_withZcs} is 0.9, indicating good compatibility between the model and the data.\par  
\begin{table}[tph]
\centering
\caption{The measured masses and widths of the $Z_{cs}(3985)^0$ and $Z_{cs}(3985)^+$~\cite{Zcs}.}
\begin{tabular}{ccc} \hline\hline
                 & Mass (MeV/$c^2$)                & Width (MeV) \\
\hline
$ Z_{cs}(3985)^0$ & $3992.2\pm{1.7}\pm{1.6}$        & $7.7_{-3.8}^{+4.1}\pm{4.3}$ \\
$ Z_{cs}(3985)^+$ & $3985.2^{+2.1}_{-2.0}\pm{1.7}$ & $13.8_{-5.2}^{+8.1}\pm{4.9}$ \\
\hline\hline
\end{tabular}
\label{tab:CompareZcs}
\end{table}
According to the fitted signal yields in Table~\ref{tab:cs_Zcs}, the Born cross section of $\ee\to \bar{K}^{0} \zcsn$ multiplied by the 
branching fraction of $\zcsn$ decays, $\sigma^{\rm Born}(\ee\rightarrow
\bar{K}^{0} Z_{cs}^{0}+c.c.)\times{\mathcal{B}}(Z_{cs}^{0}\rightarrow \Dsp\Dstm+\Dsstp\Dm)$,
can be obtained by the following equation
\begin{equation}
\sigma^{\rm Born} \times \mathcal{B} =  \frac{N^{\rm obs}}{ 2 \mathcal{L} \times \hat{\epsilon}\times(1+\delta)\times\delta_{\rm vac}},
\label{eq:cs}
\end{equation}
where $N^{\rm obs}$ are the signal yields, $\hat{\epsilon}$ are  the combined MC-determined reconstruction efficiencies in the two $D$-tag methods, $\mathcal{L}$ is the integrated luminosity, $(1+\delta)$ is the radiative correction factor, and $\delta_{\rm vac}$ is the vacuum-polarization correction factor~\cite{VP_factor}; their values are given in Table~\ref{tab:cs_Zcs}. We assume  $\mathcal{B}(\zcsn\to\Dsp\Dstm)=\mathcal{B}(\zcsn\to\Dsstp\Dm)$.
The factor of 2 in the denominator in Eq.~\eqref{eq:cs} is necessary because of the equal transition rate of ${K}^{0}$ and $\bar{K}^{0}$ to $\kshort$. The cross section results at the five center-of-mass energies are listed in Table~\ref{tab:CompareZcsCross}. The $\chi^2$ of each energy point is defined as the square of difference of the cross sections of two channels divided by the sum-of-squares of these uncertainties.
The $\chi^{2}_{\rm{total}}/\rm{ndf}$ is the sum of the $\chi^{2}$ divided by the number of energy points. 
The cross section results for the neutral channel are consistent with those for the charged one~\cite{Zcs},
which agree with the prediction based on isospin symmetry.\par
\begin{table}[tp]
\centering
\caption{Summary of the integrated luminosity ($\mathcal{L}$), the number of signal events ($N^{\rm obs}$), reconstruction
efficiency ($\hat{\epsilon}$), radiative-correction factor ($1+\delta$), and vacuum
polarization factor ($\delta_{\rm vac}$).}
\begin{tabular}{ccccc} \hline\hline
$\sqrt{s}$ (MeV) & $\mathcal{L}$(pb$^{-1}$) & $N^{\rm obs}$ & $\hat{\epsilon}~(\%)$ & $(1+\delta)\delta_{\rm vac}$ 
\\
\hline
4628 & 511.1  & $14.4^{+8.9}_{-7.5}$     & 1.88 & 0.69    \\
4641 & 541.4  &  $0.0^{+5.8}_{-0.0}$     & 1.88 & 0.74    \\
4661 & 523.6  & $10.0^{+6.9}_{-5.9}$     & 1.83 & 0.77     \\
4682 & 1643.4 &$25.5^{+13.6}_{-11.4}$    & 1.80 & 0.79    \\
4699 & 526.2  & $26.1^{+8.4}_{-7.5}$     & 1.78 & 0.80   \\
\hline\hline
\end{tabular}
\label{tab:cs_Zcs}
\end{table}
\begin{table}[tp]
\centering
\caption{Born cross sections multiplied by branching fraction of $\bar{K}^0  Z_{cs}(3985)^0$ and $\kaonm  Z_{cs}(3985)^+$ at the 5 energy points.  The $\chi^2/{\rm ndf}$ quantifies the compatibility of the five measurements.}
\begin{tabular}{ccccc} \hline\hline
\multirow{2}{*}{$\sqrt{s}$ (MeV)} &  \multicolumn{2}{c}{$\sigma^{\rm Born}\times{\mathcal{B}}~\rm{(pb)}$}  & \multirow{2}{*}{$\chi^2$} & \multirow{2}{*}{$\chi^2_{\rm total}$/ndf}\\
                             &  {\tiny $\bar{K}^0  Z_{cs}(3985)^0$} & {\tiny $\kaonm  Z_{cs}(3985)^+$} & & \\
\hline                                           
4628 & $4.4^{+2.6}_{-2.2}\pm2.0$ & $0.8^{+1.2}_{-0.8}\pm{0.6}$&1.2&\multirow{5}{*}{5.1/5}\\ 
4641 & $0.0^{+1.6}_{-0.0}\pm0.2$ & $1.6^{+1.2}_{-1.1}\pm{1.3}$&0.5&\\
4661 & $2.8^{+1.8}_{-1.6}\pm0.6$ & $1.6^{+1.3}_{-1.1}\pm{0.8}$&0.3&\\
4682 & $2.2^{+1.2}_{-1.0}\pm0.8$ & $4.4^{+0.9}_{-0.8}\pm{1.4}$&1.0&\\
4699 & $7.0^{+2.2}_{-2.0}\pm1.8$ & $2.4^{+1.1}_{-1.0}\pm{1.2}$&2.1&\\
\hline\hline
\end{tabular}
\label{tab:CompareZcsCross}
\end{table}
Systematic uncertainties on the measurement of the $\zcsn$ resonance parameters and production cross sections are extensively investigated as detailed in Ref.~\cite{supple}. An important contribution is associated with the background modelling in the fit and the $\zcsn$ signal model. For the background modelling, we vary the size and shape of the combinatorial backgrounds according to the $M(D)$ sideband control samples, as well as explore the additional contributions from the highly excited $D_{(s)}^{**}$ states. For the signal modelling, we test different $J^{P}$ assignments of the $\zcsn$ by changing the matrix elements in the signal simulations.
The total systematic uncertainties are, overall, similar to the statistical uncertainties on each measurement.

In summary, based on data sets with center-of-mass energies from 4.628~\gev~to 4.699~\gev~at BESIII, evidence of a neutral open-strange hidden-charm state, $Z_{cs}(3985)^{0}$, is found in the $\kshort$
recoil-mass spectrum of the $\ee\to\kshort(\DDsone + \DDstwo)+c.c.$ processes, with
a resonance mass and width determined as $(3992.2\pm{1.7}\pm{1.6})\mevcc$ and 
$(7.7_{-3.8}^{+4.1}\pm{4.3})\mev$, respectively.
The significance of the state is determined to be
$4.6\sigma$. 
Since this state decays through $\Dsp\Dstm$ and $\Dsstp\Dm$, it
should contain at least four quarks, $c\bar{c}\bar{s}d$.
The measured mass of the $Z_{cs}(3985)^{0}$ is 
larger than that of the $Z_{cs}(3985)^{+}$, which is consistent with theoretical prediction~\cite{Wan:2020oxt}. 
In addition, the Born cross sections of $\ee\to \bar{K}^0 Z_{cs}(3985)^{0} + c.c.$ multiplied by the branching
fraction of $Z_{cs}(3985)^{0} \to \DDsone + \DDstwo$ at the five energy points are measured and found to be consistent with those of $\ee\to K^- Z_{cs}(3985)^{+}+c.c.$~\cite{Zcs}, as is expected under isospin symmetry.
Hence, we conclude that the $Z_{cs}(3985)^{0}$ is the
isospin partner of the $Z_{cs}(3985)^+$.
\par
The BESIII collaboration thanks the staff of BEPCII and the IHEP computing center for their strong support. This work is supported in part by National Key R\&D Program of China under Contracts Nos. 2020YFA0406400, 2020YFA0406300; National Natural Science Foundation of China (NSFC) under Contracts Nos. 11521505, 11635010, 11735014, 11805086, 11822506, 11835012, 11935015, 11935016, 11935018, 11961141012, 12022510, 12025502, 12035009, 12035013; National 1000 Talents Program of China; the Chinese Academy of Sciences (CAS) Large-Scale Scientific Facility Program; Joint Large-Scale Scientific Facility Funds of the NSFC and CAS under Contract No. U1832207; CAS Key Research Program of Frontier Sciences under Contract No. QYZDJ-SSW-SLH040; 100 Talents Program of CAS; Fundamental Research Funds for the Central Universities, Lanzhou University, University of Chinese Academy of Sciences; INPAC and Shanghai Key Laboratory for Particle Physics and Cosmology; ERC under Contract No. 758462; European Union Horizon 2020 research and innovation programme under Contract No. Marie Sklodowska-Curie grant agreement No 894790; German Research Foundation DFG under Contracts Nos. 443159800, Collaborative Research Center CRC 1044, FOR 2359, GRK 2149; Istituto Nazionale di Fisica Nucleare, Italy; Ministry of Development of Turkey under Contract No. DPT2006K-120470; National Science and Technology fund; STFC (United Kingdom); The Royal Society, UK under Contracts Nos. DH140054, DH160214; The Swedish Research Council; U. S. Department of Energy under Contract No. DE-FG02-05ER41374.\par

\clearpage
\begin{center}
\textbf{\boldmath\large Supplemental Material for ``Evidence for a neutral near-threshold structure in the $K_S^0$ recoil-mass spectra
in $e^+e^-\rightarrow K_S^0D_s^+D^{*-}$ and $e^+e^-\rightarrow K_S^0D_s^{*+}D^{-}$"}
\end{center}
\normalsize
\appendix
\section{Fit results of $D_s$-tag and $D$-tag methods}
\begin{figure}[h]
\centering
\includegraphics[width=0.6\textwidth]{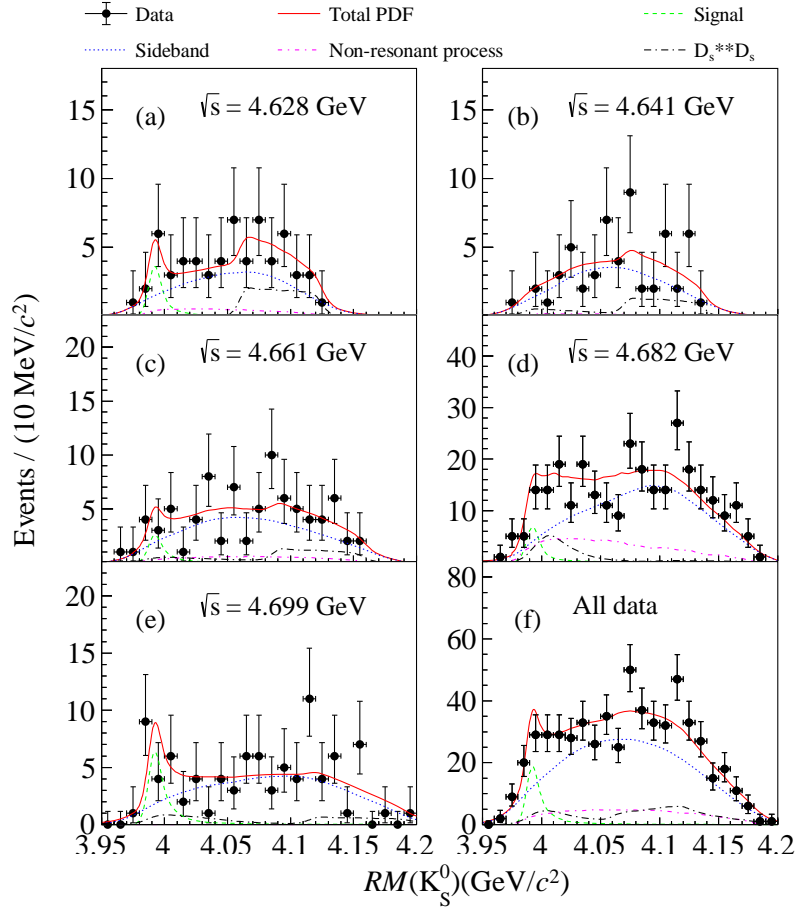}
\caption{Fits to the recoil mass distributions $RM(K_S^0)$ using $D_s^+$-tag method.}
\label{fig:M_recKs_Ds_appendix}
\end{figure}

\begin{figure}[h]
\centering
\includegraphics[width=0.6\textwidth]{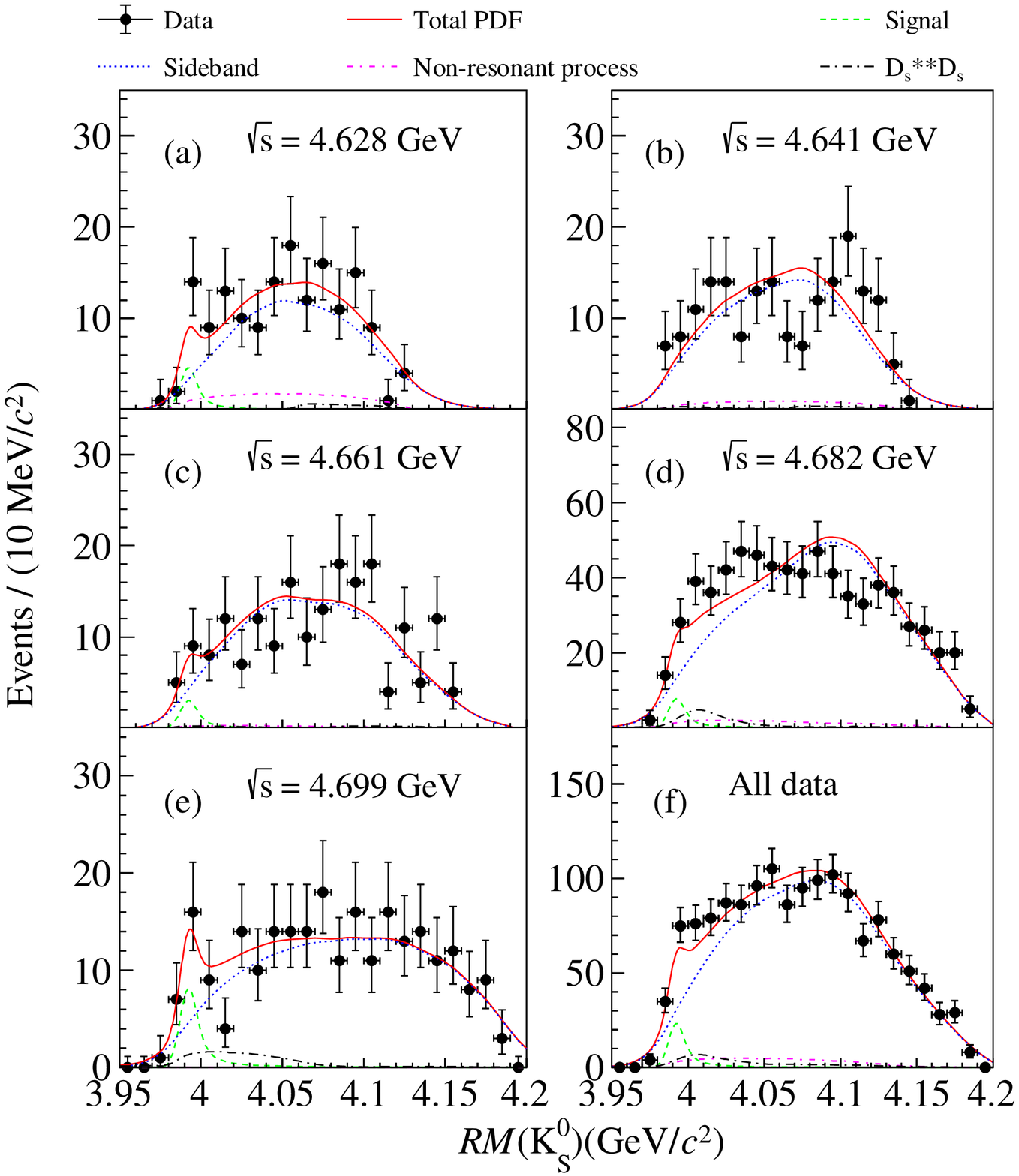}
\caption{Fits to the recoil mass distributions $RM(K_S^0)$ using $D^-$-tag method.}
\label{fig:M_recKs_D_appendix}
\end{figure}

Figure~\ref{fig:M_recKs_Ds_appendix} and Fig.~\ref{fig:M_recKs_D_appendix} show the fits to the recoil mass distributions $RM(K_S^0)$ using only the $D_s^+$-tag or $D^-$-tag method, respectively. Table~\ref{tab:M_recKs_DDs} lists the yields of the two methods.
\begin{table}[htp]
\centering
\caption{Yields of $K_{S}^{0}Z_{cs}^{0}$ with $D_s^+$-tag and $D^-$-tag methods.}
\renewcommand{\arraystretch}{1.3}
\begin{tabular}{ccc}\hline\hline
$\sqrt{s}$(MeV)  & $\Dsp$-tag & $\Dm$-tag\\
\hline                                           
4628 &  $6.5^{+4.1}_{-3.4}$ &  $7.8^{+4.9}_{-4.1}$\\
4641 &  $0.0^{+2.7}_{-2.3}$ &  $0.0^{+3.1}_{-2.7}$\\
4661 &  $4.6^{+3.2}_{-2.7}$ &  $5.3^{+3.7}_{-3.1}$\\
4682 & $12.0^{+6.4}_{-5.3}$ & $13.6^{+7.2}_{-6.1}$\\
4699 & $12.2^{+3.9}_{-3.5}$ & $14.0^{+4.5}_{-4.0}$\\
\hline\hline
\end{tabular}
\label{tab:M_recKs_DDs}
\end{table}
\section{Fit results based on two subsets of data sample at $\sqrt{s}=4.682~\rm{GeV}$}
To avoid potential bias, the analysis strategy is firstly implemented using 1/3 of the data set at $\sqrt{s}=4.682~\rm{GeV}$. The remaining 2/3 of data set at $\sqrt{s}=4.682~\rm{GeV}$ is then used for a consistency check.   \par
\begin{figure}[h]
\centering
\includegraphics[width=0.45\textwidth]{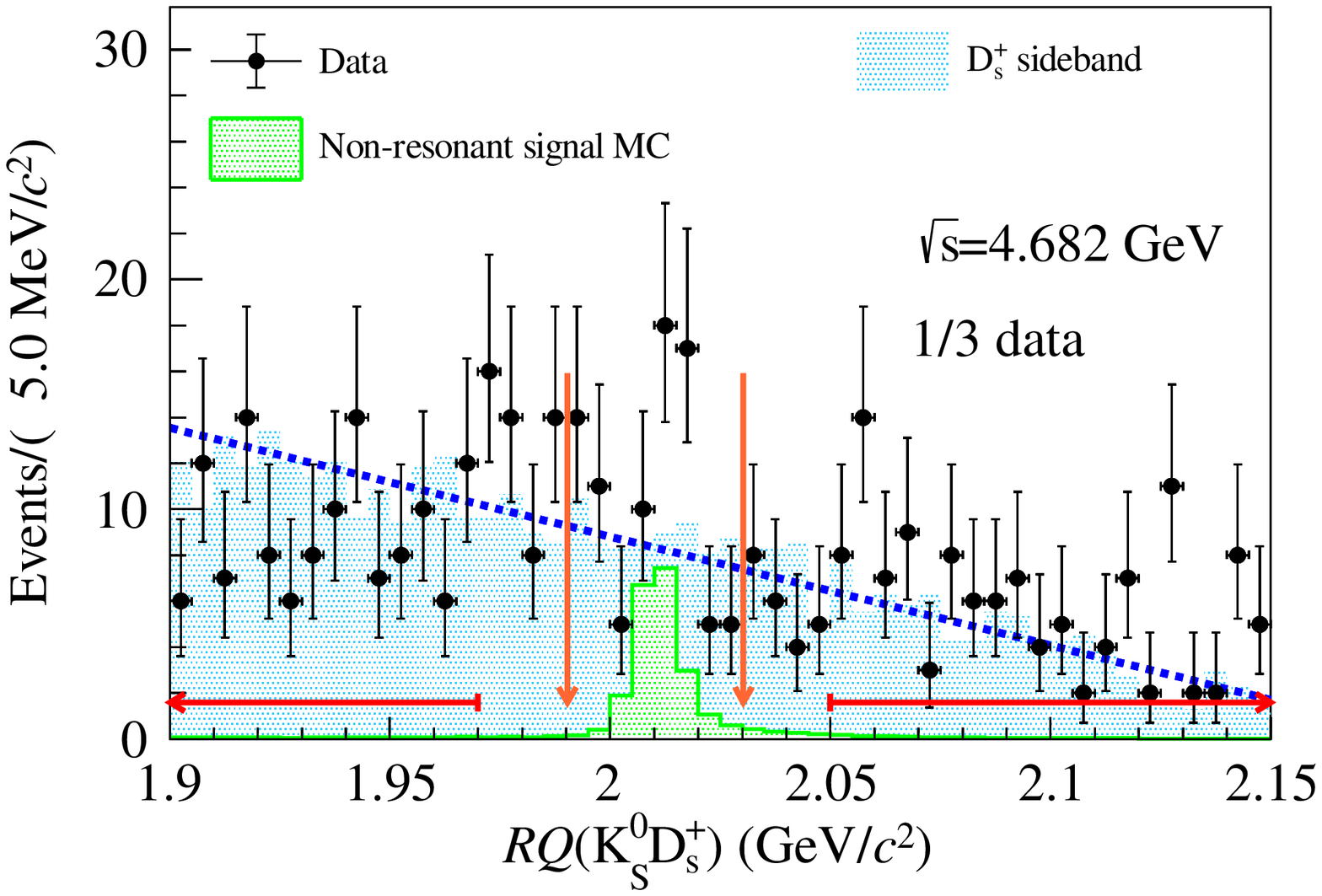}
\includegraphics[width=0.45\textwidth]{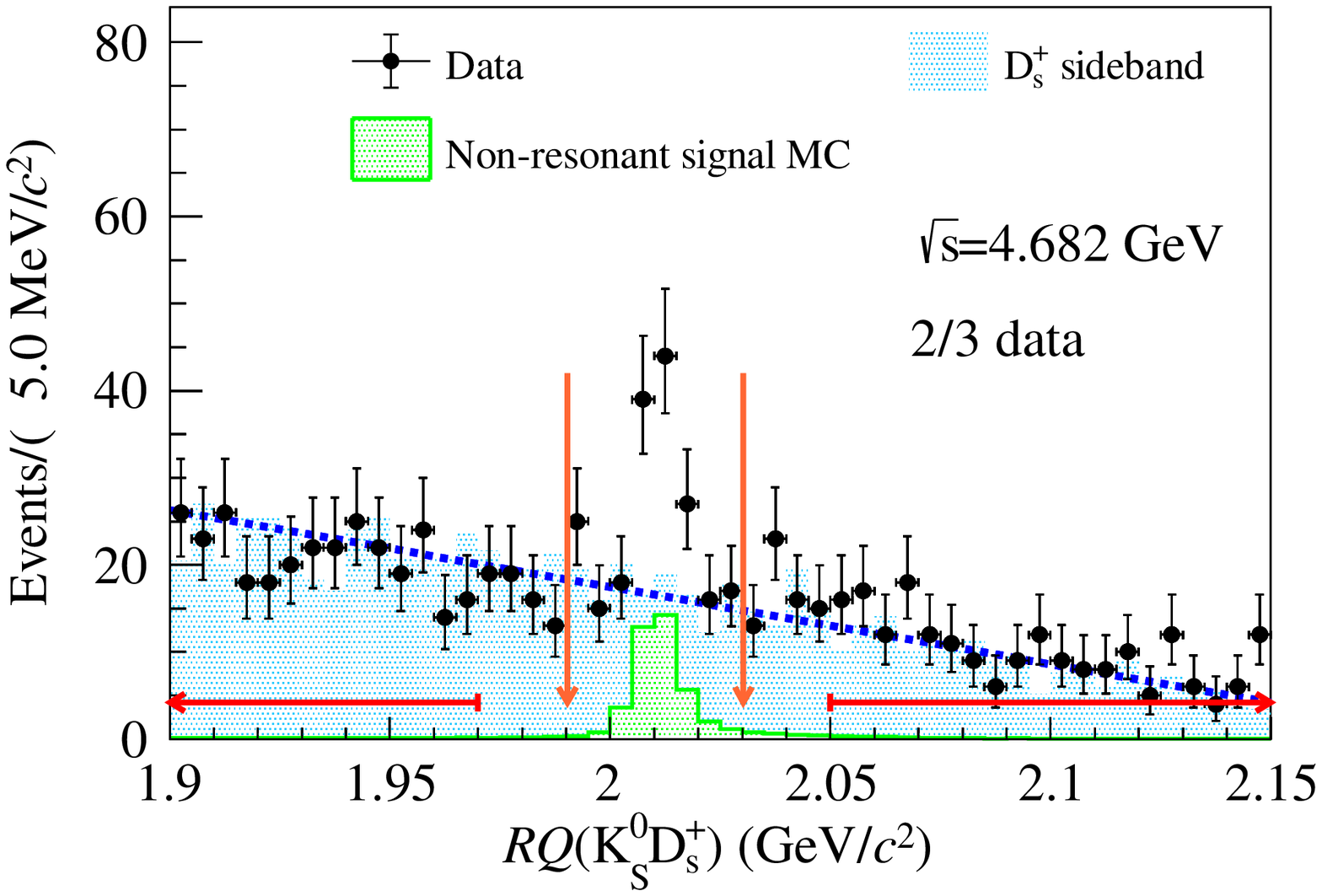}\\
\caption{Fits to the recoil mass distributions $RQ(K_S^0D_s^{+})$ from two data subsets at 4.682 GeV.}
\label{fig:M_recKsDs_appendix}
\end{figure}
\begin{figure}[h]
\centering
\includegraphics[width=0.45\textwidth]{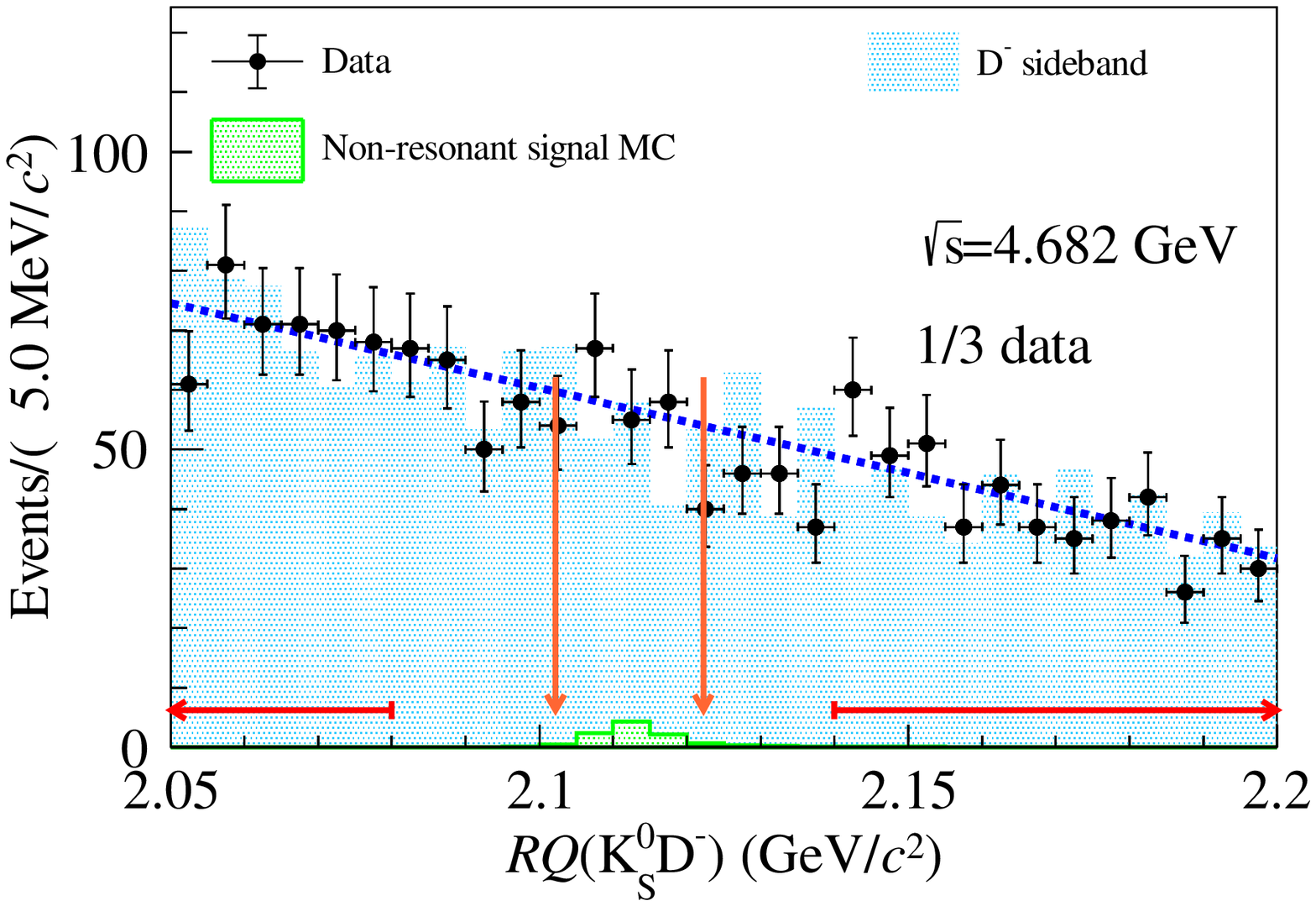}
\includegraphics[width=0.45\textwidth]{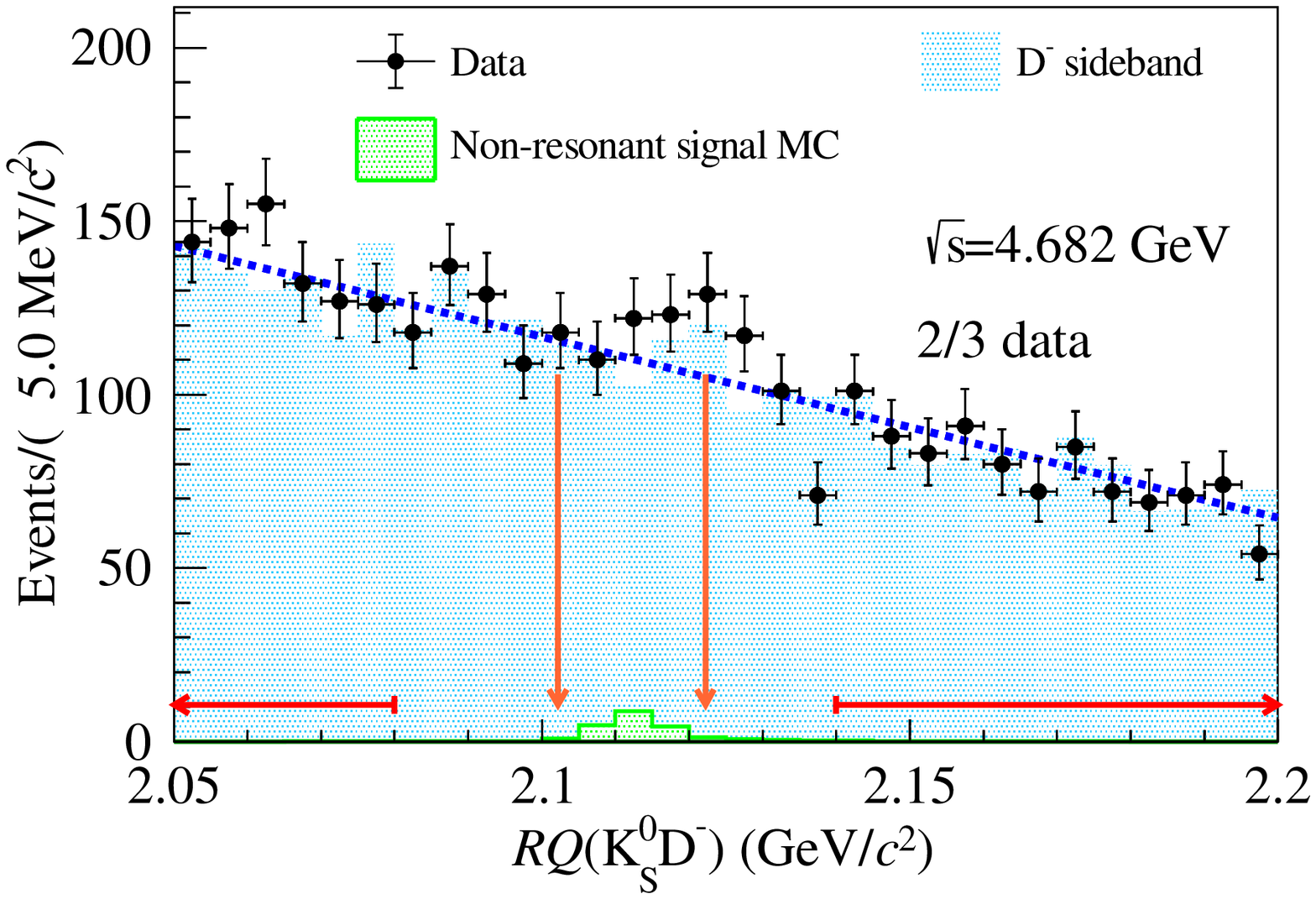}\\
\caption{Fits to the recoil mass distributions $RQ(K_S^0D^{-})$
from  two data subsets at 4.682 GeV.}
\label{fig:M_recKsD_appendix}
\end{figure}

The distributions of $RQ(K_S^0D_s^+)$ and $RQ(K_S^0D^-)$ 
are shown in Fig.~\ref{fig:M_recKsDs_appendix} and 
Fig.~\ref{fig:M_recKsD_appendix}, respectively, for the two subsets.
Using the method described in the paper, we determine the number of combinatorial background events in the   
$D^{*-}$ and $D_{s}^{*+}$ signal regions that are listed in Table~\ref{tab:combbkg_appendix}. The ratios between the numbers from two data subsets are all consistent with two, as expected.\par
\begin{figure}[h]
\centering
\includegraphics[width=0.45\textwidth]{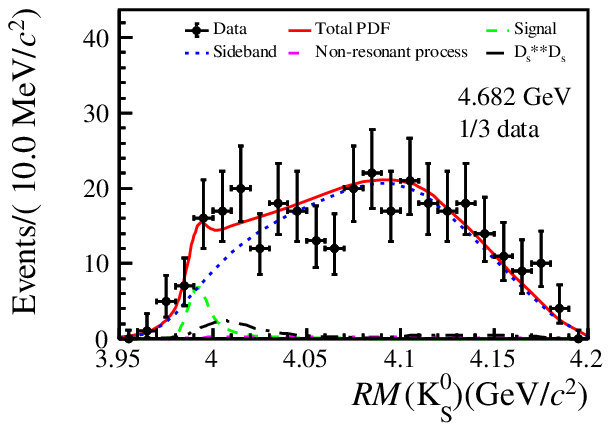}
\includegraphics[width=0.45\textwidth]{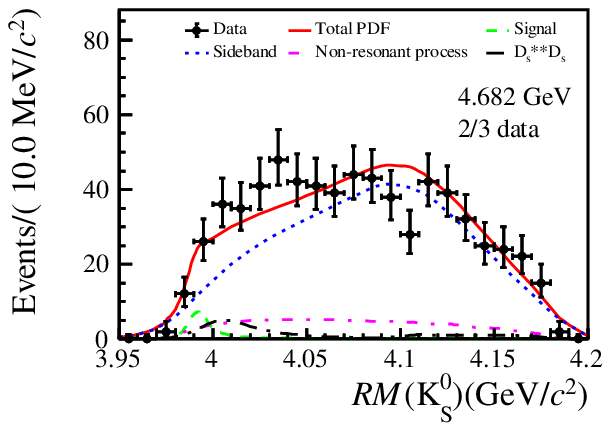}
\caption{Fit to the recoil mass distribution $RM(K_S^0)$ from  two data subsets at 4.682 GeV.}
\label{fig:fit_withZcs_appendix}
\end{figure}

The $RM(\kshort)$ distributions  from two data subsets
are shown in Fig.~\ref{fig:fit_withZcs_appendix}.
The yields of $Z_{cs}(3985)^{0}$ are obtained from the fits to the distributions, which are listed in Table~\ref{tab:sig_appendix}. 
The ratio between the number of $Z_{cs}(3985)^{0}$ signal events from two data subsets is also consistent with two.   \par
\begin{table}[htbp]
\centering
\caption{Number of combinatorial backgrounds events in the signal regions of $\kshort\Dsp\Dstm$ and $\kshort\Dsstp\Dm$ three-body processes.}
\begin{tabular}{c|c|c} \hline
                         & Tag $\Dsp$       & Tag $\Dm$                \\
\hline                                           
1/3 of 4682 data & $  66.6\pm4.2$& $227.4\pm 7.6$ \\
2/3 of 4682 data & $132.4\pm5.9$& $441.5\pm10.5$ \\
\hline
Ratio               & $2.0\pm0.2$     & $1.9\pm0.1$  \\
\hline
\end{tabular}
\label{tab:combbkg_appendix}
\end{table}
\begin{table}[htbp]
\centering
\caption{Yields of the $Z_{cs}(3985)^{0}$ from two data subsets at 4.682 GeV.}
\begin{tabular}{c|c} \hline
                             & Yield of $Z_{cs}(3985)^{0}$         \\
\hline                                           
1/3 of 4682 data & $12.8^{+8.4}_{-6.9}$\\
2/3 of 4682 data & $12.7^{+9.8}_{-8.3}$\\
\hline
Ratio                    & $1.0\pm1.0$ \\
\hline
\end{tabular}
\label{tab:sig_appendix}
\end{table}
\clearpage

\section{Two-dimensional distributions}
\begin{figure}[htbp]
\centering
\includegraphics[width=0.7\textwidth]{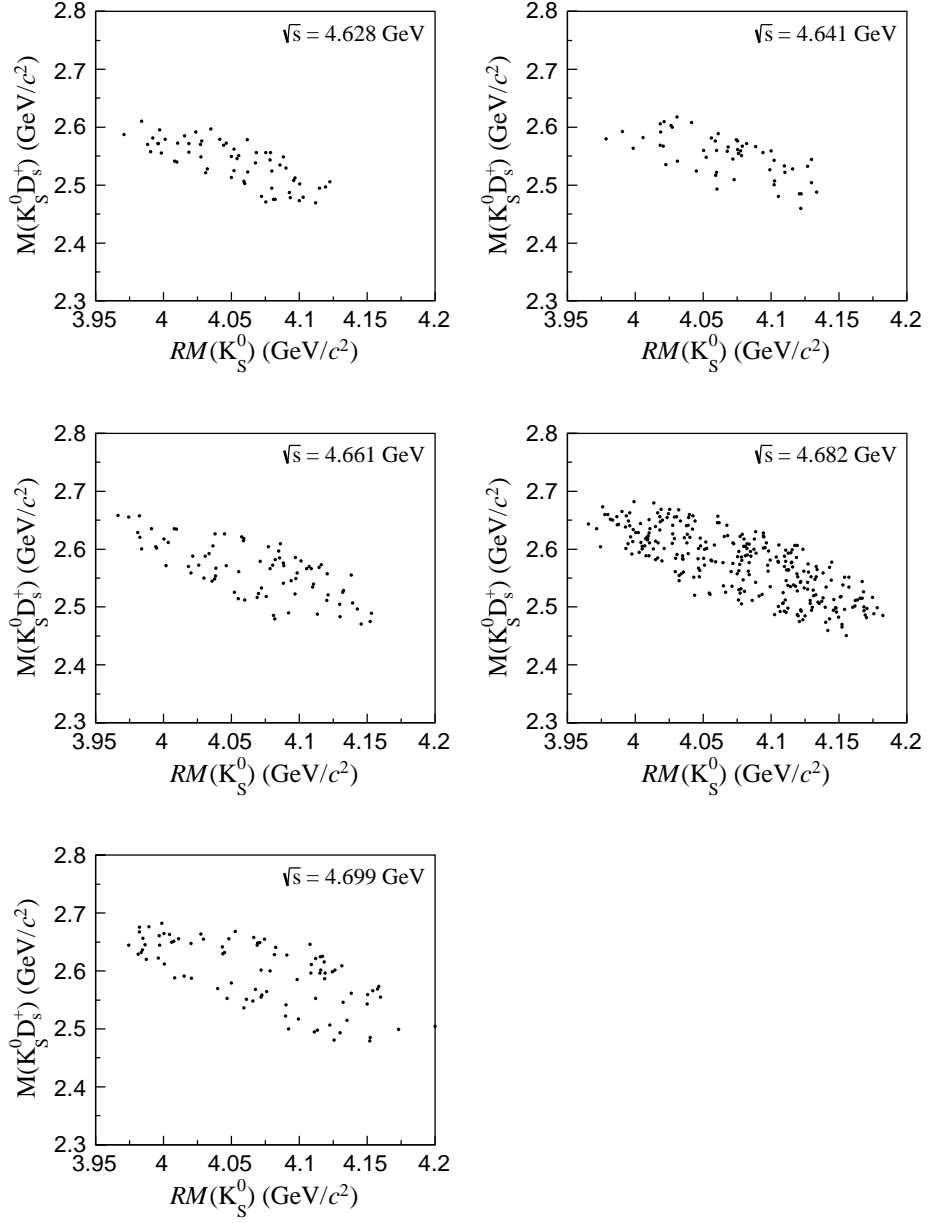}\\
\caption{
Two-dimensional distributions of  $M(K_{S}^{0}D_s^+)$ vs $RM(K_S^0)$ in data.
}
\label{fig:2D_DsTag}
\end{figure}

\begin{figure}[htbp]
\centering
\includegraphics[width=0.7\textwidth]{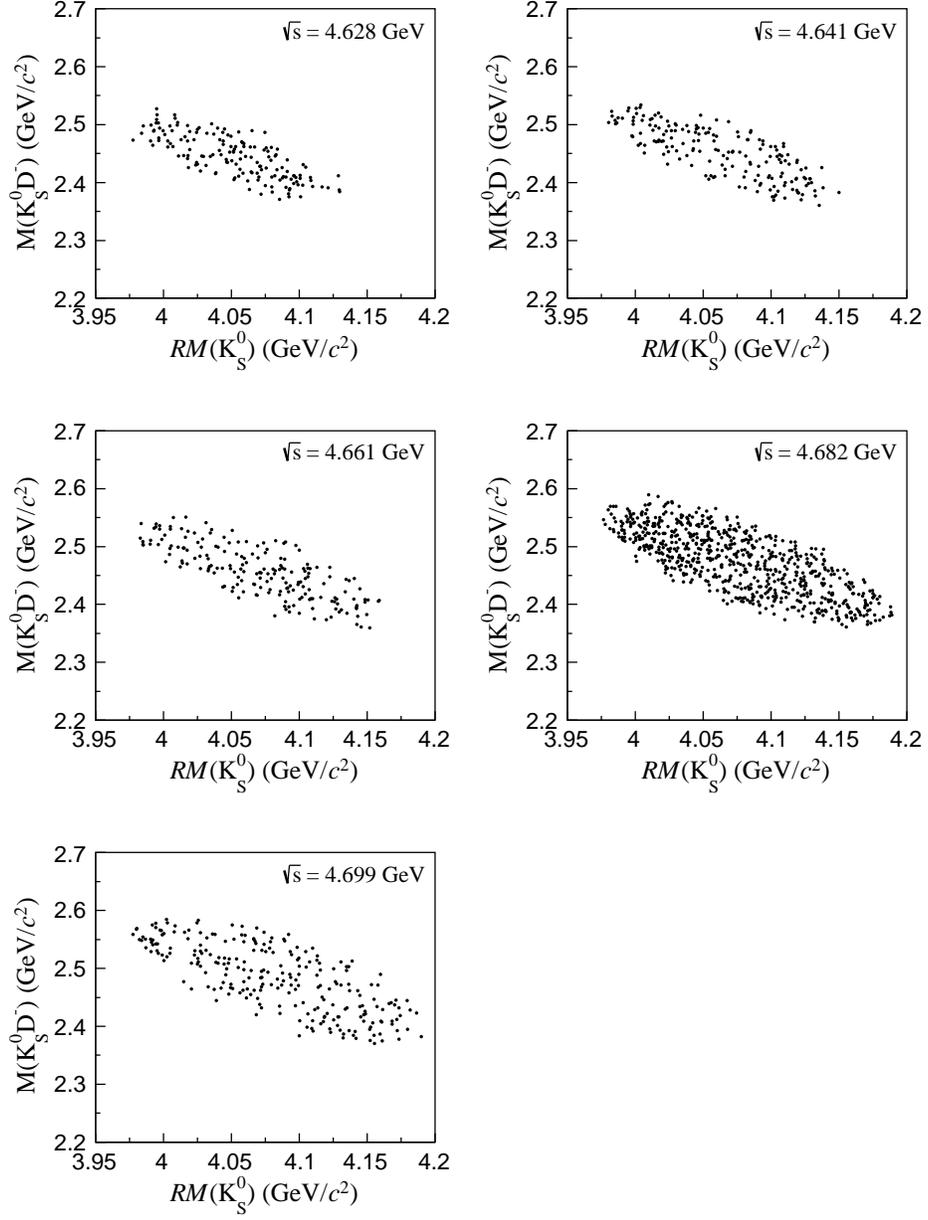}\\
\caption{
Two-dimensional distributions of $M(K_{S}^{0}D^-)$ vs $RM(K_S^0)$ in data.
}
\label{fig:2D_DTag}
\end{figure}

Figures~\ref{fig:2D_DsTag} and \ref{fig:2D_DTag} show the two-dimensional distributions of  $M(K_{S}^{0}D^-)$ vs $RM(K_S^0)$, and $M(K_{S}^{0}D_s^+)$ vs $RM(K_S^0)$ in data.
\clearpage

\section{Systematic studies}
The total systematic uncertainties on the $Z_{cs}(3985)^0$ resonance parameters
and cross sections are the quadrature sums of the assigned uncertainties arising from the sources discussed below.  A summary of these contributions is listed in Table~\ref{tab:sys}. When including all these sources of systematic uncertainty, the significance of the 
$Z_{cs}(3985)^0$ signal becomes $4.6\sigma$.\par

In the nominal fit to the $RM(K_S^0)$ spectra, we choose to constrain the width of the $\zcsn$ with the uncertainty of the  $Z_{cs}^+$ width, to improve the precision of our measurement, according to the isospin symmetry of the $\zcsn$ and $Z_{cs}^+$. If the constraint is removed, the fit width of the $\zcsn$ becomes $4.1^{+4.7}_{-3.9}$ (stat. only), which is consistent with the nominal result. 

\label{sec:syst}
\begin{table*}[htbp]
  \begin{center}
   \caption{Summary of systematic uncertainties on the $\zcsn$ resonance parameters and cross section. Also given are the resultant significances when including each individual contribution together with the statistical uncertainty. ``$\cdots$" means the uncertainty is negligible.}
  \resizebox{\linewidth}{!}{
  \begin{tabular}{l|c|c|c|c|c|c|c|c}
 \hline \hline
 Source   & Mass(MeV/$c^2$)  & Width(Mev)  &
 $\sigma^{\rm born}_{4.628}\cdot\mathcal{B}$(pb) &
 $\sigma^{\rm born}_{4.641}\cdot\mathcal{B}$(pb) &
 $\sigma^{\rm born}_{4.661}\cdot\mathcal{B}$(pb) &
 $\sigma^{\rm born}_{4.682}\cdot\mathcal{B}$(pb) &
 $\sigma^{\rm born}_{4.699}\cdot\mathcal{B}$(pb) & $Z_{cs}^0$ Significance \\ \hline
 Tracking                &     &     &  2.8$\%$ &  2.8$\%$ &  2.8$\%$  & 2.8$\%$ & 2.8$\%$ &\\
 Particle ID             &     &     &  2.8$\%$ &  2.8$\%$ &  2.8$\%$  & 2.8$\%$ & 2.8$\%$ &\\
 $\kshort$               &     &     &  2.4$\%$ &  2.4$\%$ &  2.4$\%$  & 2.4$\%$ & 2.4$\%$ &\\
 $\pi^0$,$\eta$          &     &     &  0.1$\%$ &  0.1$\%$ &  0.1$\%$  & 0.1$\%$ & 0.1$\%$ &\\
$D^{-}$/$D_{s}^{+}$ signal window && &  0.5$\%$ &  0.5$\%$ &  0.5$\%$  & 0.5$\%$ & 0.5$\%$ &\\
 Mass  scale             & 0.8 &     &      &      &       &     &     &\\
 Resolution              & 0.1 & 0.1 &  1.2$\%$ &  1.2$\%$ &  1.2$\%$  & 0.9$\%$ & 0.8$\%$ &$5.0\sigma$\\
 $f$ factor              & 0.8 & 0.4 &  4.9$\%$ &  7.2$\%$ &  9.5$\%$  & 6.7$\%$ &14.4$\%$ &$4.8\sigma$\\
 Signal model            & 0.5 & 3.1 &  9.8$\%$ &  8.6$\%$ &  7.3$\%$  & 6.1$\%$ &15.2$\%$ &$5.3\sigma$\\
 Backgrounds             & 0.4 & 2.2 & 35.9$\%$ & 27.2$\%$ & 18.5$\%$  &12.7$\%$ & 6.8$\%$ &$4.9\sigma$\\
 Efficiency              & 0.1 & 0.1 &  0.7$\%$ &  0.5$\%$ &  0.2$\%$  & 1.0$\%$ & 0.6$\%$ &$5.0\sigma$\\
 $D_{(s)}^{**}$ states   & 0.6 & 1.6 & 19.8$\%$ & 14.5$\%$ &  9.2$\%$  &26.6$\%$ & 9.4$\%$ &$4.6\sigma$\\
 $\sigma^{\rm born}(\kshort\zcsn)$& 0.6 & 1.1 & 12.1$\%$ &  6.8$\%$ &  1.5$\%$  & 7.0$\%$ & 1.8$\%$ &$4.7\sigma$\\
 Luminosity              &      &     & 1.0$\%$ &  1.0$\%$ &  1.0$\%$  & 1.0$\%$ & 1.0$\%$ & \\
 Input BFs               &      &     & 2.8$\%$ &  2.8$\%$ &  2.8$\%$  & 2.8$\%$ & 2.8$\%$ & \\
     \hline
     total & 1.6 & 4.3 & 44.5$\%$ & 34.0$\%$ & 24.6$\%$ & 32.1$\%$ & 24.7$\%$ \\
   \hline\hline
    \end{tabular}
}
    \label{tab:sys}
  \end{center}
  \end{table*}
 {\bf Tracking, PID and reconstruction of intermediate states:}
    The uncertainties on both the tracking and PID efficiencies for each charged track are assigned to
    be 1\%.  The uncertainties associated with $\kshort$,
    $\pi^0$ and $\eta$ reconstruction are assigned to be
    2\%. The uncertainties from tracking, PID and
    intermediate states reconstruction in different tag channels are weighted by
    the factor $\mathcal{B}_{l}\varepsilon_{l}$.  Here, ``$l$" indicates each
    $D^+$ or $D_s^-$ decay channel. 

 {\bf $D^{-}$/$D_{s}^{+}$ signal window:}
    The uncertaintiy associated with the defintion of the $D^{-}$/$D_{s}^{+}$ signal window is estimated by comparing
    the $D^{-}$/$D_{s}^{+}$ signal from data and MC. The widths of the $D^{-}$/$D_{s}^{+}$ peaks in data and MC are slightly
    different. We estimate that these differences in resolution lead to a relative $0.5\%$ difference in efficiency, which is assigned as a systematic uncertainty.

 {\bf Mass scale:}
    A control sample of
    $\ee\to\kshort\Dm\Dsp$ events with $\sqrt{s}$ larger than $4.62\gev$ is selected,
    in which the $\kshort$ and $\Dsp$ are reconstructed. We fit the $\Dm$ peak in the corrected recoil mass spectrum      $RM(\kshort\Dsp)+M(\Dsp)-m(\Dsp)$. 
    The $\Dm$ signal is modelled with a MC-determined signal shape convolved with a Gaussian
    function. The Gaussian parameters are
    determined to be $\mu$=$(0.07\pm 0.68)\mevcc$ and $\sigma$=$(0.60\pm2.62)\mev$.
    Since the corrected recoil mass $RM(\kshort\Dsp)+M(\Dsp)-m(\Dsp)$ is largely insensitive to the resolution of the $\Dsp$ mass, we attribute any mass shift to the bachelor $\kshort$.
    Hence, cconsidering the central value and uncertainty of this study, we take a maximum mass shift of $0.8\mevcc$ as the 
    systematic uncertainty.

 {\bf Detector resolution:}
    To understand the potential difference of detector resolution in data and MC
    simulations, the same control sample of $\ee\to\kshort\Dm\Dsp$ events is used. From
    the ``Mass scale'' study , the width of the smearing function is  at most $3.2\mev$. We therefore smear
    the resolution function in the $\zcsn$ fit by this amount and reperform the mass fit. The
    resultant differences on the final results 
    are taken as systematic uncertainties.

 {\bf{$f$ factor:}}
    In the default fit, the two signal processes $\zcsn\to\Dsp\Dstm$ and
    $\zcsn\to\Dsstp\Dm$ are combined  and we assume their fraction factor is
    $0.5$ in nominal calculation. To estimate the possible systematic bias arising from this source, we assume
    the probability distribution of $f$ is uniform between 0 and 1 with no prior
    knowledge, we take the RMS value of 1/$\sqrt{12}$ (~0.3) as the uncertainty on $f$.
    Hence, we vary $f$ to 0.2 and 0.8 and take the
    largest difference with respect to the nominal result as the systematic uncertainty
    from this source.

 {\bf Signal model:} 
    In the default fit, we assume the $J^{P}$ of the $\zcsn$ is $1^{+}$ and that the 
    $\kshort$ and $\zcsn$ in the rest frame of the $\ee$ system and
    the $\Dsp$($\Dsstp$) and $\Dstm$($\Dm$) in the $\zcsn$
    system are both in an $S$-wave state, denoted as ($S$, $S$). As a systematic check, we 
    also consider $0^{-}$($P$, $P$), $1^{-}$($P$, $P$), $1^{+}$($D$, $S$) and $2^{-}$($P$,$P$) configurations. 
    To minimize the effect of systematic uncertainties in the study, 1000 toy MC samples are generated with the PDF
    determined from the default fit of $RM (K_{S}^{0})$. The number of events in each sample is the same as  the  data sample.     The whole analysis procedure is repeated under 
    different $J^{P}$ assumptions and the mean fit result from the ensemble of toys is measured.
    We take the largest difference with respect to the $1^{+}$($S$, $S$) configuration as being the systematic uncertainty associated with the signal model.

 {\bf Backgrounds:} 
    The signal description is sensitive to  the contribution from background. We vary the number of 
    combinatorial background within $1\sigma$. We also use the
    inclusive MC samples  instead of  the $D^-(D_{s}^+)$ sideband samples to
    extract the shape of combinatorial backgrounds. We take the largest difference with respect to 
    the default result as the sytematic uncertainty.

 {\bf Efficiency functions:} 
    We vary the parameters of the efficiency functions used in the fitting to estimate the
    impact of the modelling of the acceptance. The parameters of the efficiency functions are
    varied within 1$\sigma$, and the new functions are used to refit.
    The relative differences with the nominal results
    are taken into account as systematic uncertainties. 

 {\bf Highly excited $D_{(s)}^{**}$ states:}    
    In the default analysis, we include the contributions of three highly excited
    $D_{s}^{**}$ processes, $D^{*}_{s1}(2536)^{-}(\to\Dstm\kshort)\Dsp$,
    $D^{*}_{s2}(2573)^{-}(\to\Dm\kshort)\Dsstp$ and
    $D^{*}_{s1}(2700)^{-}(\to\Dstm\kshort)\Dsp$, based on the results  of the control samples
    studied in the charged $Z_{cs}(3985)^+$ analysis~\cite{Zcs}. Another
    potential $D^{**}$ background is $D^{*}_{1}(2600)^{+}(\to D_{s}^{+}\kshort)D^{*-}$.
    According to the study of the charged $Z_{cs}(3985)^+$~\cite{Zcs}, the ratio 
    $\mathcal{B}(D_{1}^{*}(2600)^{0}\rightarrow D_{s}^{+}K^{-})/\mathcal{B}(D_{1}^{*}(2600)^{0}\rightarrow D^{+}\pi^{-})=0.00\pm0.02$.
    Assuming the cross section 
    of $D^{*}_{1}(2600)^{+}D^{*-}$ is the same as $D^{*}_{1}(2600)^{0}\bar{D}^{*0}$, we fix the ratio 
    $\mathcal{B}(D_{1}^{*}(2600)^{+}\rightarrow D_{s}^{+}K_{S}^{0})/\mathcal{B}(D_{1}^{*}(2600)^{+}\rightarrow D^{+}\pi^{0})$ to 0.02 in the fit. We also replace the PHSP component with other possible processes with $D_{(s)}^{**}$ states,  such as 
    $D^{*}_{3}(2750)^{+}(\to\Dsstp\kshort)\Dm$, 
    $D^{*}_{2}(2460)^{+}(\to\Dsp\kshort)\Dm$,
    $D(2550)^{+}(\to\Dsstp\kshort)\Dm$,
    $D^{*}_{1}(2600)^{+}(\to\Dsp\kshort)\Dm$, and 
    $D(2740)^{+}(\to\Dsstp\kshort)\Dm$.
    The resultant changes are assigned as the systematic uncertainty associated with this source.

 {\bf $\sigma(\kshort\zcsn)$ line shape:}
    In the detault fit, the lineshape of the $\kshort\zcsn$ cross section is
    extracted with a 4th order polynomial function, and then inserted into the  KKMC generator
    to evaluate the ISR effect in MC generation, which affects the radiative
    correction factor, detection efficiency and detection-resolution function of the 
    default result. For the systematic uncertainty study, we vary the cross
    sections within 1$\sigma$ and refit the
    lineshape. Then the signal MC samples are generated based on the new lineshape. The resultant maximum changes are taken as the 
    systematic uncertainty from this source.

 {\bf Luminosity:}
    The uncertainty of the luminosity measurement at each energy point is assigned to be
    1\%.

 {\bf Branching fractions:}
    In this analysis, the branching fractions of $\kshort\to\pip\pim$, $\Dstm\to\Dm
    X$ and all the decay channels used in the  $\Dsp$ and $\Dm$ reconstruction are taken from the PDG~\cite{pdg}. 
   We use the quoted uncertainties on these quantities to determine the corresponding
   systematic uncertainties for our measurements.

\end{document}